\documentclass[pra,twocolumn,showpacs,preprintnumbers,amsmath,amssymb,tightenlines,epsfig]{revtex4}

\usepackage{graphicx}% Include figure files
\usepackage{dcolumn}% Align table columns on decimal point
\usepackage{bm}% bold math
\usepackage{epsfig}
\usepackage{stmaryrd}

\newcommand{\vc}[1]{\mathbf{#1}}
\newcommand{\figuresize}{\columnwidth}
\newcommand{\smallfiguresize}{4.2cm}

\newcommand{\bv}[1]{\mbox{\boldmath$ #1 $}}

\def\beq{\begin{equation}}
\def\eeq{\end{equation}}
\def\nnl{\\[0.15cm] \nonumber}

\def\CR{\nonumber\\[0.15cm]}
\newcommand{\fref}[1]{Fig.~\ref{#1}}
\newcommand{\eref}[1]{Eq.~(\ref{#1})}
\newcommand{\sref}[1]{section \ref{#1}}
\begin{document}

\title{Numerical Study of the stability of Skyrmions in Bose-Einstein Condensates}
\author{S. W\"uster, T.E. Argue and C.M. Savage}
\affiliation{ARC Centre of Excellence for Quantum-Atom Optics, Department of Physics, 
Australian National University, Canberra
ACT 0200, Australia }
\email{sebastian.wuester@anu.edu.au}

\begin{abstract}
We show that the stability of three-dimensional Skyrmions in trapped Bose-Einstein condensates depends critically on scattering lengths, atom numbers, trap rotation and trap anisotropy.  In particular, for the $^{87}$Rb  $|F=1,m_{f}=-1\rangle$, $|F=2,m_{f}=1\rangle$ hyperfine states stability is sensitive to the scattering lengths at the $2\%$ level, where the differences between them are crucial. In a cigar shaped trap, we find stable Skyrmions with slightly more than 
$2\times10^{6}$ atoms, a number which scales with the inverse square root of the trap frequency. These can be stabilized against drift out of the trap by laser pinning.
\end{abstract}

%03.75.Lm - topological excitations in BECs, 03.75.Mn - multi-component BECs

\maketitle

\section{Introduction}

A broad range of physics has been uncovered in the study of vortices in Bose-Einstein condensates (BECs) \cite{fetter:review,Kevrekidis:review,stringari:review}. Therefore adding internal degrees of freedom \cite{book:pethik} to produce multi-component topological defects, such as coreless vortices \cite{ketterle:coreless} and Skyrmions, should reveal further interesting phenomena.
Skyrmions, which are the subject of this paper, were first introduced in particle physics \cite{skyrme:no1,skyrme:no2,witten:no1,witten:no2}, where their application \cite{diakonov:pentaquarks} has recently triggered a controversial search for penta-quarks \cite{experiment:pentaquarks}. 

It has been shown that Skyrmions can be energetically stable in a harmonically trapped two component BEC \cite{savage:skyrm}, and that the Skyrmion topology can be seeded by phase imprinting techniques \cite{ruoste:imprint1,ruoste:imprint2,ruoste:imprint3}. However their experimental realization remains to be achieved. In order to facilitate this, we have surveyed the relevant parameter space and identified regimes in which Skyrmions are stable. 

Other work on Skyrmions in BECs includes their study in a homogeneous background \cite{battye:homog} and in spin-1 spinor BECs \cite{stoof:spinor}. Trapped Skyrmions with winding numbers greater than one have also been shown to be stable \cite{ruoste:highw}.
\begin{figure}
\centering
\epsfig{file={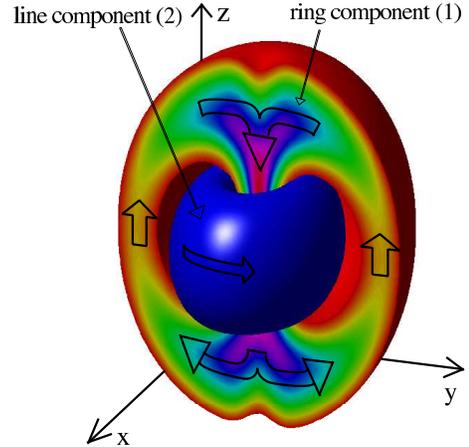},width=6cm}
\caption{(color online) Three dimensional density profiles for a Skyrmion with $N_{tot}=2\times10^{6}$, $N_{r}=0.6$ in the presence of a pinning potential ($V_{0}=6.5\hbar \omega_{x}$, $w_{0}=1.3\sigma_{x}$, see \eref{laserpinning}) and in a rotating trap ($\Omega=0.06\omega_{x}$). The anisotropy of the trap is $\omega_{\perp}/\omega_{z}=1.3$ and the scattering lengths are the $\bar{a}_{ij}$. See text for definitions of symbols and for values of other parameters. The central (blue) torus is an isosurface of the line component. An isosurface of the ring component (red) is shown for $x < 0$: on the $y$-$z$ plane between the isosurface sections the ring component density is indicated by a colormap from red (lowest) to purple (highest). The circulation associated with each vortex is also indicated. }
\label{3dplot}
\end{figure}

Skyrmions are topological solitons. Their two component wavefunction has a three dimensional structure that cannot be continuously deformed into a simpler structure, such as a coreless vortex. A defining property of the Skyrmion is that the atomic field approaches a constant value at spatial infinity: for a trapped BEC this means sufficiently far from the trap centre. Then we can identify all points at infinity, turning the domain of our wave function into the compact space $S^{3}$, the three dimensional spherical surface in four dimensions. 
The two component wavefunction of the BEC at a particular position can be represented by the $SU(2)$ matrix which maps a standard state, such as $(1,0)$, into it. The wavefunction is then represented by a map from $S^{3}$ onto $SU(2)$. A fundamental result of homotopy theory is that all such mappings can be classified by an integer topological invariant, the winding number $W$ \cite{rajaraman:solitons}. 

For a $W=1$ Skyrmion, one component of the BEC carries a charge one line vortex, while the other component circulates around the first and through the line vortex core, carrying a charge one ring vortex singularity (\fref{3dplot}). The outer component carrying a ring singularity will be called the ``ring component'' (index 1), and the inner component with a line vortex will be called the ``line component'' (index 2). The filling of the ring vortex core by the line component stabilizes the dual vortex structure as outlined in \cite{savage:skyrm}.

Considering the specific case of $^{87}$Rb  hyperfine states $|1\rangle \equiv |F=1,m_{f}=-1\rangle$, $|2\rangle \equiv |F=2,m_{f}=1\rangle$, and identical spherical traps with frequency $\omega=2 \pi \times7.8$ Hz, we found a lower limit on the atom number for stability against cylindrically symmetric perturbations of about $3 \times 10^{6}$ atoms in a spherical trap. This limit is slightly lower if the trap is elongated. 
Stabilization against general perturbations, which requires trap rotation, is only possible with at least $9 \times 10^{6}$ atoms. 
However with laser pinning, and slight adjustment of the scattering lengths, Skyrmions with as few as $2 \times 10^{6}$ atoms are found to be stable. The scaling relation given in the next section allows these results to be applied to traps with other frequencies. 

A result with particular significance for the experimental realization of Skyrmions is that their stability depends on the atom-atom interaction strengths at the $2\%$ level, where the differences between them are crucial.
% - which is less than the current level of experimental uncertainty of $2\%$.

This paper is organized as follows: in \sref{model} we introduce the underlying physics and relevant parameters. Section \ref{Numbers} presents the results of our search for low atom number Skyrmions, section \ref{offset} contains a brief discussion of  trap offset between the two components, \sref{drift} shows how to stabilize Skyrmions against drift, and \sref{scattlength} gives an overview of their sensitivity to scattering lengths. The appendix discusses the origin of the trap offset between the components.

\section{Method}
\label{model}
To determine Skyrmion stability we numerically solve the Gross-Pitaevskii equation (GPE) in imaginary time, which corresponds to a numerical minimisation of the energy functional (see e.g. \cite{garcia:numerics}).  The coupled GPEs for a two component BEC in a frame rotating with angular velocity $\mathbf{\Omega}$ compared to the lab frame are
\begin{align}
i\hbar \frac{\partial}{\partial t}\Psi_{i}(\vc{x})&=\Big(-\frac{\hbar^{2}}{2m}\nabla^{2} + V(\vc{x}) + \sum_{j=1,2} U_{ij} \left| \Psi_{j}(\vc{x})\right|^{2}
\CR
&-\mathbf{\Omega} \cdot \mathbf{L} \Big)\Psi_{i}(\vc{x}) ,
\label{GPEs}
\end{align}
where $\Psi_{i}$ for $i=1,2$ denote the BEC wavefunctions of the two components, normalised to the number of atoms in each component, $N_{1}$ and $N_{2}$. For later reference we define $N_{tot}=N_{1}+N_{2}$ and $N_{r}=N_{2}/ N_{tot} $. $V(\vc{x})= m (\omega_{\perp}^{2}(x^{2}+y^{2})+\omega_{z}^{2}z^{2})/2$ is the harmonic trapping potential, and $m$ is the atomic mass. The interaction parameters $U_{ij}$ are proportional to the scattering lengths for intra- $(U_{11},U_{22})$ and inter-component ($U_{12}$) interactions: $U_{ij}=4\pi \hbar^{2} a_{ij}/m$. 
$\mathbf{L}$ is the angular momentum operator.

The imaginary time GPEs follow from \eref{GPEs} by the replacement $\tau=it$. To solve them, we start with a trial wavefunction that is in the $W=1$ topological class  \cite{savage:skyrm}. During the imaginary time evolution the normalisation of the wavefunctions is held fixed at $N_{1}$, $N_{2}$. If the wavefunctions converge, and are stable against general perturbations, the result is a local minimum of the energy. This is a steepest descent method in the function space with fixed norm, whereby the initial trial function is evolved in the direction where the energy decreases most rapidly. 

Experimental Skyrmion creation would involve phase imprinting a state in the $W=1$ topological class onto the BEC \cite{ruoste:imprint1,ruoste:imprint2,ruoste:imprint3}. This would then relax towards the ground state of the class by means of thermal dissipation. This process can be simulated numerically using a combined real and imaginary time formalism \cite{ballagh:vortexgen}.

%%%%%%%%%%%%%%%%%%%%
After rescaling, $\tilde{\vc{x}}=\vc{x}/a_{z}$, $\tilde{t}=\omega_{z}t$, $\tilde{\Psi}_{i}=\Psi_{i}a_{z}^{3/2}/\sqrt{N_{i}}$, $\tilde{\mathbf{\Omega} }=\mathbf{\Omega}/\omega_{z}$, $\tilde{\mathbf{L}}=\mathbf{L}/\hbar$, where $a_z = \sqrt{\hbar / m \omega_z}$ is the harmonic oscillator scale length,
\eref{GPEs} has  the dimensionless form:
\begin{align}
i \frac{\partial}{\partial \tilde{t}}\tilde{\Psi}_{i}(\tilde{\vc{x}})&=\Big(-\frac{1}{2}\tilde{\nabla}^{2} 
+ \frac{1}{2} \left(   \frac{  \omega_{\perp}^{2}  }{  \omega_{z}^{2}  } (\tilde{x}^{2} +\tilde{y}^{2}) + \tilde{z}^{2}   \right)
\CR
&
+ \sum_{j=1,2} \lambda_{ij} \left| \tilde{\Psi}_{j}(\tilde{\vc{x}})\right|^{2}
-\tilde{\mathbf{\Omega}} \cdot \tilde{\mathbf{L}} \Big)\tilde{\Psi}_{i}(\tilde{\vc{x}}),
\label{dimlessGPEs}
\end{align}
where the interaction coefficients are defined as
\begin{align}
\lambda_{ij}&=4\pi N_{j}a_{ij} \sqrt{m \omega_{z}/\hbar}.
\label{lambdas}
\end{align}
In a system with a spherical nonrotating trap the physics is fully determined by the $\lambda_{ij}$. All simulations used the parameters of $^{87}$Rb. Early measurements gave the scattering lengths: $\bar{a}_{11}= 5.67 \mbox{ nm}, \bar{a}_{22}=5.34 \mbox{ nm}, \bar{a}_{12}=\bar{a}_{21}=5.5 \mbox{ nm}$ \cite{hall:scattlength} for the hyperfine states $ |1\rangle$ and  $|2\rangle$. 
After the bulk of our numerical work was done, we became aware of revised  values for the scattering lengths: $a_{11}=5.315 \mbox{ nm}, a_{22}=5.052 \mbox{ nm}, a_{12}=a_{21}=5.191 \mbox{ nm}$ \cite{cornell:scattlength,verhaar:scattlength}. 
These differ from the $\bar{a}_{ij}$ primarily by an overall scale factor. Choosing this to be $\bar{a}_{11}/a_{11} = 1.067$, according to \eref{lambdas} this is equivalent to an increase in atom numbers by the same factor, or by about $7\%$. The residual differences, although small, are significant. Therefore, our major results are presented for both sets of scattering lengths: early $(\bar{a}_{ij})$ and revised $(a_{ij})$.

In magnetic traps the two states will in general have slightly different trap frequencies due to the differential Zeeman effect \cite{cornell:scattlength,rabi}, which is reviewed in the Appendix. Taking gravity into account, this results in a small offset of the trap centres for each component which, for weak traps, can become comparable to the Skyrmion length scale $a_{z} \propto \omega_{z}^{-1/2}$. We report in \sref{offset} that this effect should and can be minimized and therefore, unless otherwise stated, we assume identical spherical traps with $\omega=\omega_{z}=\omega_{\perp}=2 \pi \times7.8$ Hz for both hyperfine components. 
In contrast, the change in non-linear interaction coefficients due to the actual different trap frequencies themselves is negligible.
%As will be shown in \sref{offset}, Skyrmions are very sensitive to any offset of the trap centres. We therefore discuss in \sref{offset} possibilities to minimize this effect.
%Unless otherwise stated, we assumed identical spherical traps with $\omega=\omega_{z}=\omega_{\perp}=2 \pi \times7.8$ Hz for both components. 

Although we present our results for specific trap frequencies, the scaling \eref{lambdas} implies that our results apply for different trap frequencies provided the atom number is scaled to keep the $\lambda_{ij}$ constant. For example if the trap frequencies are increased by a factor of four, the atom number decreases by a factor of two. This scaling applies provided the Gross-Pitaevskii equation is valid, and hence fails if the density becomes too high.

Skyrmions require that the interaction parameters $U_{ij}$ are such that the two components spatially separate. For a homogeneous system it can be shown that this corresponds to \cite{ho:phasesep,battye:phasesep}
\begin{align}
U^{2}_{12} > U_{11}U_{22} .
\label{phasesep}
\end{align}
%
%However for the atomic states we consider this is in fact is slightly violated, while it holds for the newer set of scattering lengths.

In our numerical survey of the BEC parameter space three dimensional spatial grids with $64^{3}$ grid points were used. However the specific results presented in this paper were verified on bigger grids. To solve the GPE in imaginary time we used both the RK4IP algorithm \cite{RK4IP} and the corresponding first order interaction picture method. We found that the lower order method was faster. Typical imaginary timesteps were $\Delta \tau =10^{-5}$ s.  The simulations were done with the aid of the high level programming language XMDS \cite{XMDS}.

\begin{figure}
\centering
\epsfig{file={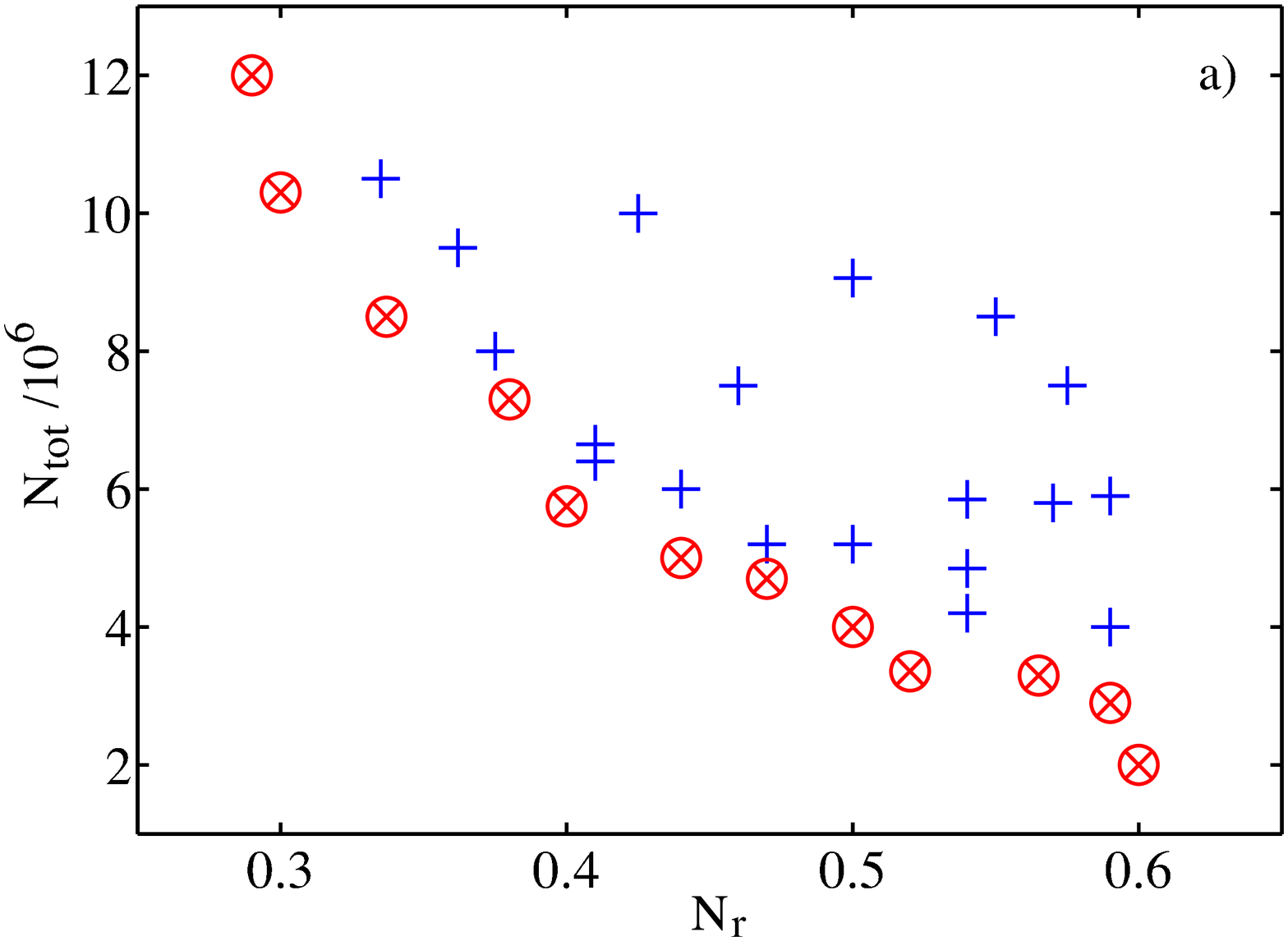},width=\figuresize}
\\
\epsfig{file={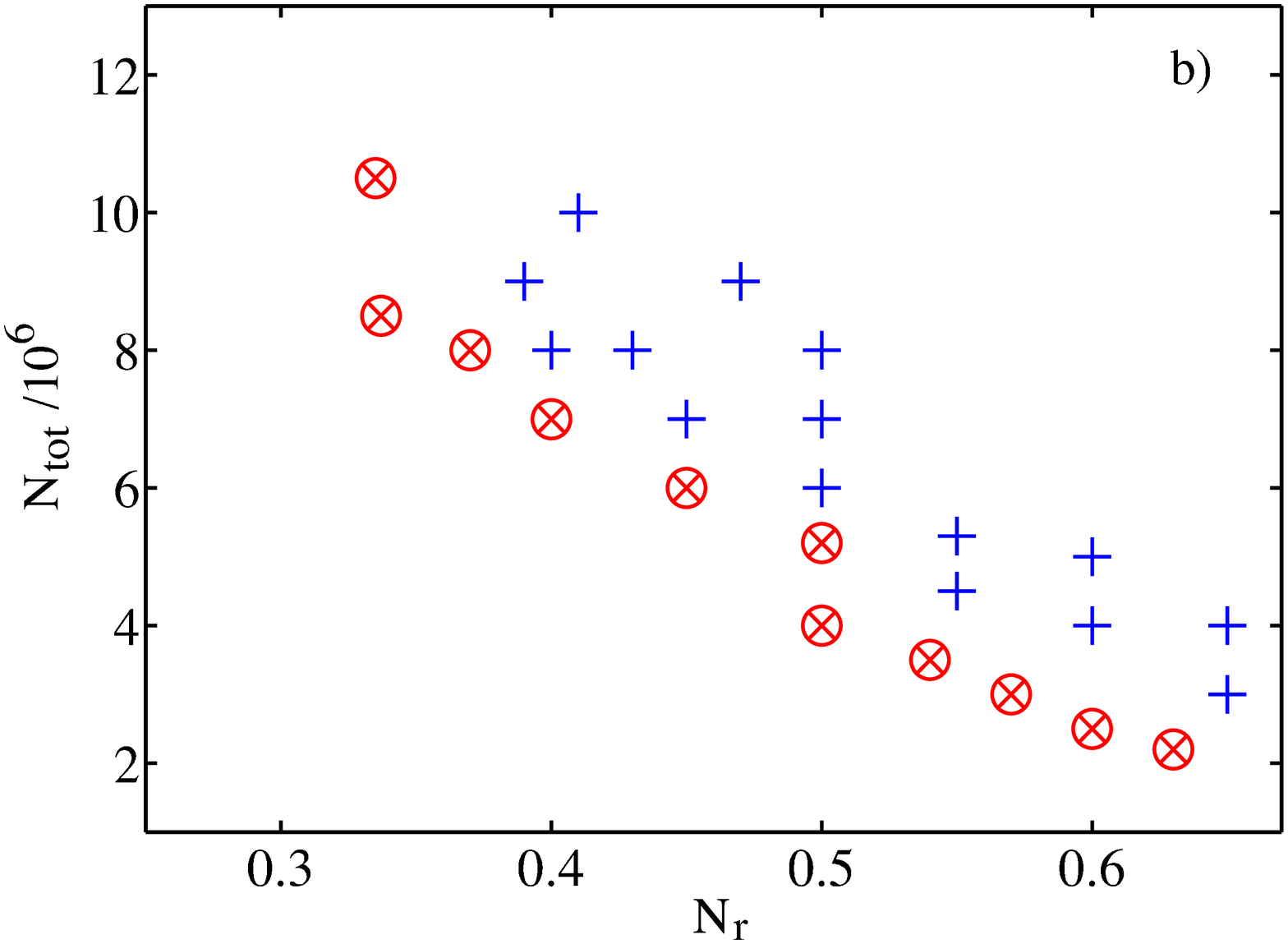},width=\figuresize}
\caption{(color online) Skyrmion stability diagram against cylindrically symmetric perturbations, in a spherical trap: ($+$ blue) stable, ($\varotimes$ red) unstable. The axes are  the total atom number $N_{tot}=N_{1}+N_{2}$ and relative line component fraction $N_{r}=N_{2}/N_{tot}$. 
a) $\bar{a}_{ij}$ set of scattering lengths. For $N_{r}>0.6$ all tested cases were unstable. 
b) $a_{ij}$ set of scattering lengths. For $N_{r}>0.65$ all tested cases were unstable.}
\label{lineofstability}
\end{figure}

\section{Stability against cylindrically symmetric perturbations}
\subsection{Atom numbers}
\label{Numbers}

Previous work had suggested that stable Skyrmions require total atom numbers of the order of $10^{7}$  \cite{savage:skyrm}.  A major motivation for the work reported in this paper was to find stable Skyrmions  with fewer atoms. We analyze stability in two stages. In the first stage, reported in this section, we consider stability against cylindrically symmetric perturbations. In particular, against collapse of the ring vortex, or deconfinement of the line component. In the latter case, the density of the line component drops off more slowly than the ring component at the edge of the condensate. Consequently the phase variation associated with the line vortex extends to infinity and the wavefunction there cannot be identified with a single point in $SU(2)$. Thus the topological classification of wavefunctions into classes with integer winding number fails. Numerically this is indicated by the calculated winding number not being an integer. In the second stage we consider stability against non-cylindrically symmetric perturbations. In particular, against drift of the line vortex singularity out of the trap.

\begin{figure}
\centering
\epsfig{file={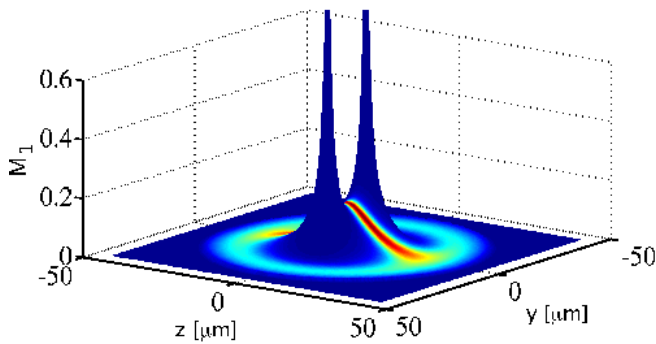},width=\figuresize}
\caption{(color online) Mach number of ring component in the y-z plane for a case near the stability boundary with $N_{tot}=5 \times 10^{6}$ and $N_{r}=0.5$. The superimposed colormap shows the ring component density (blue - lowest density, red - highest density). The peaks are at the location of the ring singularity. Scattering lengths $\bar{a}_{ij}.$}
\label{velomountain}
\end{figure}

The results of our survey of stability against cylindrically symmetric perturbations are summarized in \fref{lineofstability}. Points on the plots indicate whether a Skyrmion with the corresponding $N_{r}, N_{tot}$ values is stable ($+$) or not ($\varotimes$). The two panels show the results for the two different sets of scattering length introduced in \sref{model}. 
As expected from the discussion in that section, the revised scattering lengths require slightly more atoms for stability with given $N_r$.
%All data points were verified with simulations using $128^{3}$ grid points.
The unstable cases with slightly too few atoms have imaginary time evolution for which the ring singularity (compare with \fref{velomountain}) moves inwards until it reaches the Skyrmion core (defined as the volume with ring component flow through the line vortex core), and the flow velocity through the core increases. Eventually, the flow velocity exceeds the local two component speed of sound (see \eref{speedofsound}) and the ring vortex collapses, destroying the ring cirulation. 
%{\it In real time simulations where we increased the flow velocity through the core we also saw a breakdown of the ring circulation when it exceeded the speed of sound, which was accompanied by phonon emission \cite{landau:superfluidity}. In these cases the ring vortex was not sufficiently stabilized against its natural tendency to contract to a point \cite{grant:ring}.}

The filling of the vortex-ring by the line component can hinder the collapse of the ring vortex, and indeed we find stable Skyrmions with smaller $N_{tot}$ if the relative fraction of the filling $N_{r}$ is increased, as shown in \fref{lineofstability}. However, this works only up to  $N_{r}\sim0.65$. For higher $N_{r}$ the  line component escapes confinement by the ring component. Its phase variation means that the topological equivalence of $\mathbb{R}^{3}$ and $S^{3}$ breaks down, and with it the integer valuedness of the winding number $W$.

In summary, \fref{lineofstability} shows a region with stability  bounded towards lower $N_{tot}$ by ring vortex collapse, and towards higher $N_{r}$ by breakdown of the global topological structure. The lowest Skyrmion atom number allowed by these two constraints is $N_{tot} \sim 3 \times 10^{6}$ for the revised scattering lengths.
The increase in central flow speed associated with the collapse of the ring vortex, in imaginary time evolution, led us to consider configurations in which this velocity was reduced. The flow speed is high through the core because it is narrow, its width being determined by the healing length,  \fref{velomountain}.  The return flow speed of the ring component through the outer shell of the Skyrmion is typically orders of magnitude slower because it occurs through a correspondingly greater cross sectional area.

The quantisation of the circulation for a charge one ring vortex requires the velocity $\mathbf{v}_{1}$ to satisfy
\begin{align}
\oint_{\cal C} \mathbf{v}_{1}  \cdot \mathbf{dl}&=\frac{h}{m},
\label{flowquant}
\end{align}
where $\cal C$ is any closed path around the ring singularity. Due to the high flow speeds through the core, most of the circulation accumulates there. Elongating the core, by stretching the trap into a cigar shape, should therefore slow the core flow as the circulation is generated over a greater length of core. This is indeed what we found. However decreasing the atom numbers for which stable Skyrmions exist by stretching the trap reaches a limitation as the outer shell of the ring component thins and the line component pushes through to the outside. 

To prevent this we decreased  $a_{22}$ by $2\%$. Achieving this might require scattering length manipulation, perhaps using an optical Feshbach resonance \cite{denschlag:feshbach}. This increases the surface tension, as will be described in \sref{scattlength}, and simultaneously makes the line component denser and more localized. With this we found a stable Skyrmion with $N_{tot}=2.13 \times 10^{6}$ atoms and $N_{r}=0.6$ in a trap with anisotropy $\eta \equiv \omega_{\perp}/\omega_{z}=1.3$, $\omega_{\perp} = 2 \pi \times7.8$ Hz.

\begin{figure}
\centering
\epsfig{file={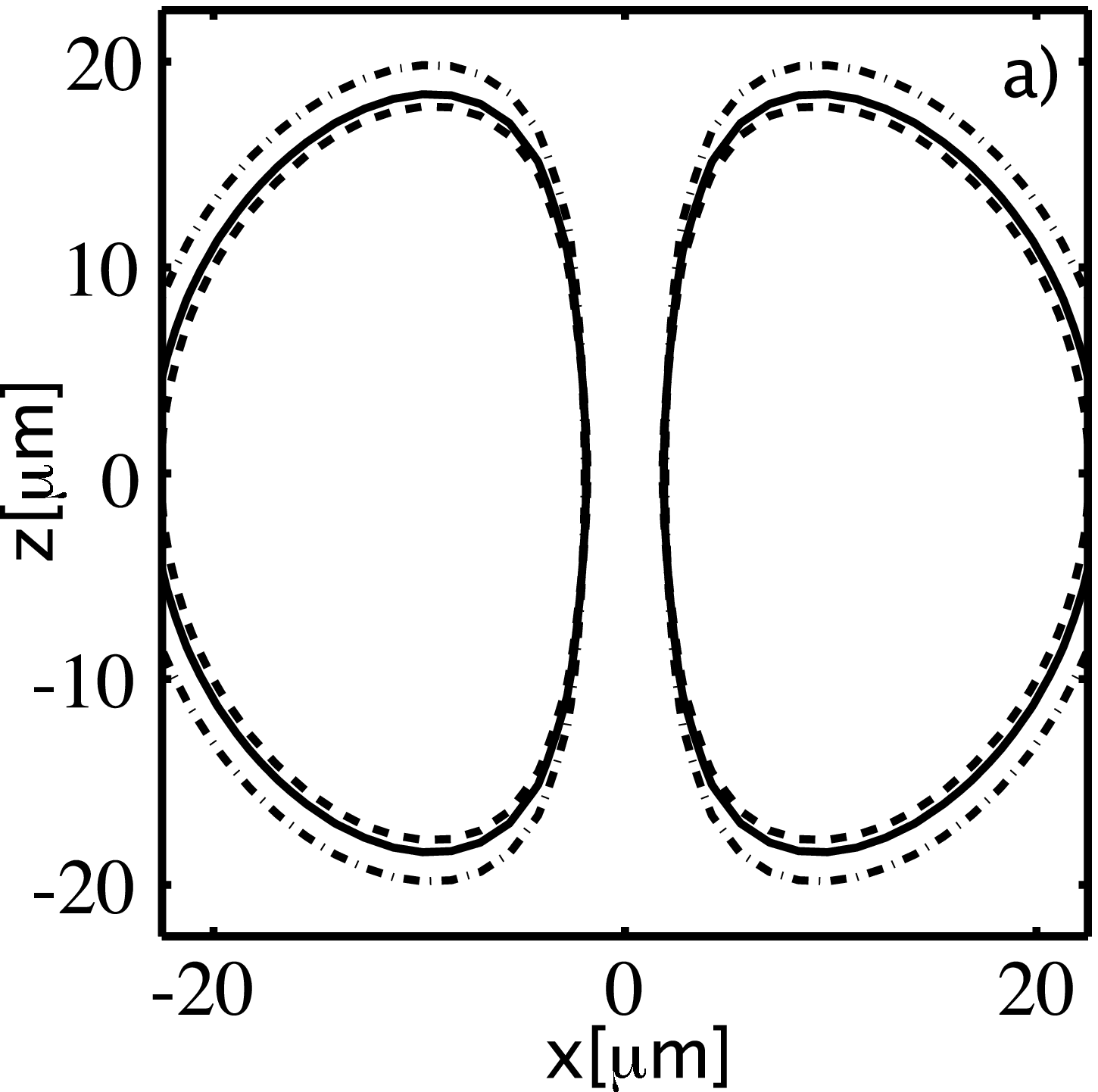},width=5.8cm}
\\
\epsfig{file={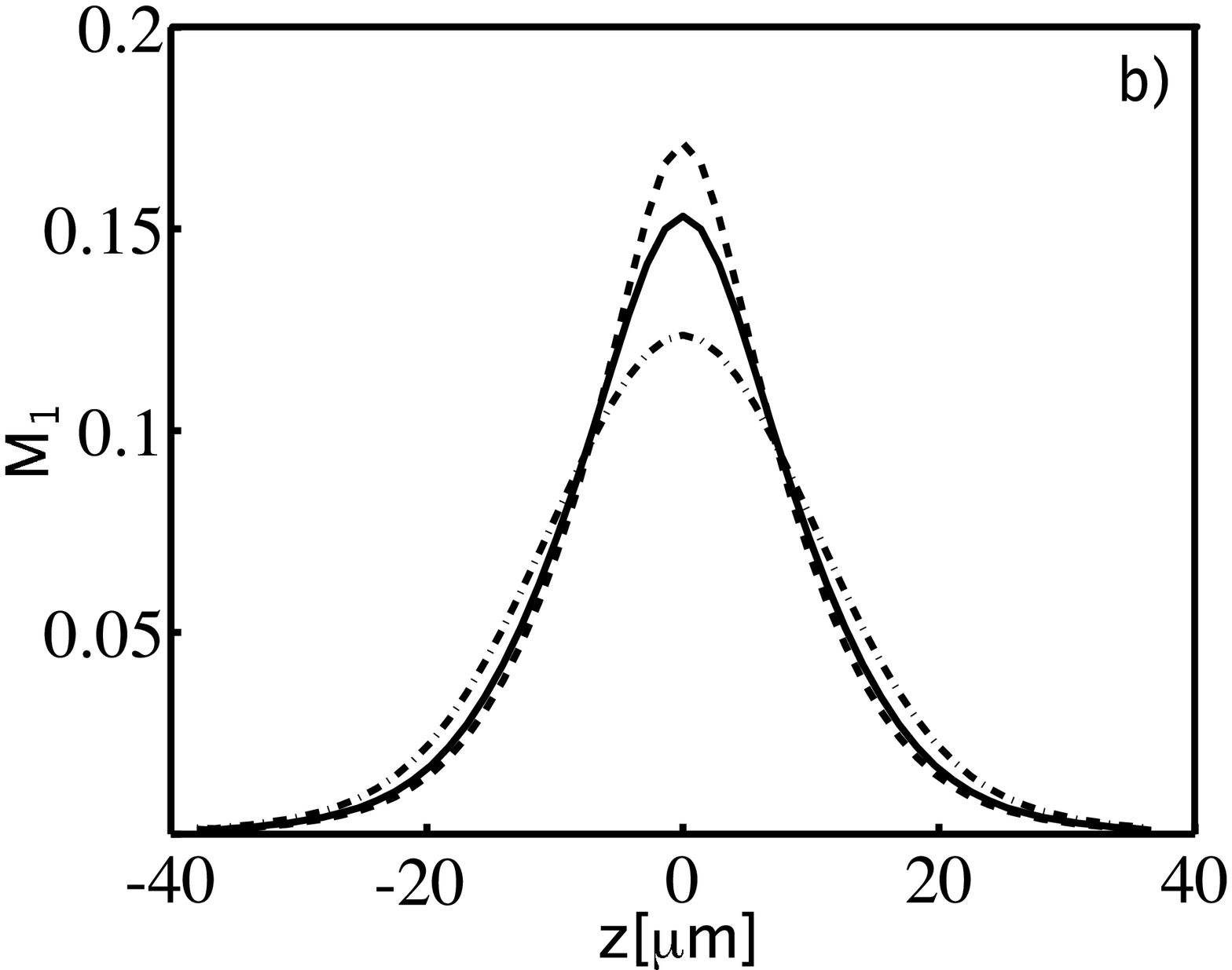},width=6cm}
\caption{Influence of changes in $N_{r}$ on the core ($N_{tot}=5 \times 10^{6}$): (Dashed) $N_{r}=0.47$, (solid) $N_{r}=0.5$, (dot-dashed) $N_{r}=0.58$. Scattering lengths $\bar{a}_{ij}$. a) Shape of the core. Shown is the cross section of an isosurface of the density of the line component ($n_{2}=2.5\times10^{-19}\mbox{m}^{-3}$) with the x-z plane. b) Mach number of the ring component $M_{1}$ along the core for x=0, y=0.}
\label{stretching}
\end{figure}

We also observed that increasing the fraction of line component $N_{r}$, that has the stabilizing effect described above, lowers the central flow speed: \fref{stretching} shows how the shape of the core varies for small deviations in $N_{r}$ and the effect of this variation on the flow speed.

\begin{figure}
\centering
\epsfig{file={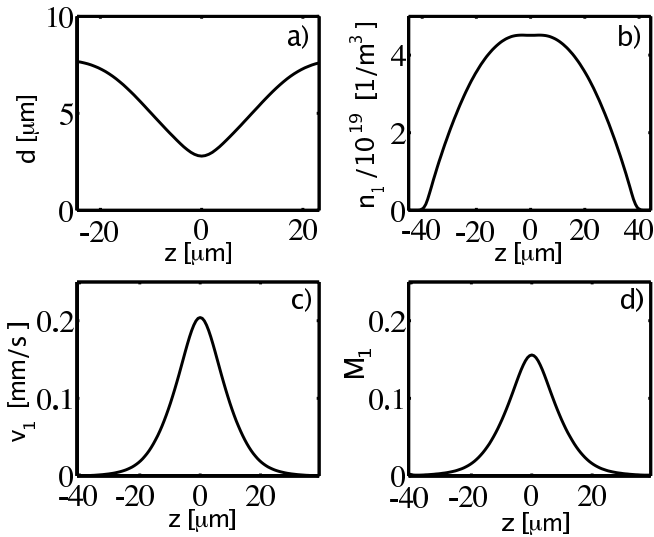},width=\figuresize}
\caption{Skyrmion core properties ($N_{tot}=5 \times 10^{6}$ and $N_{r}=0.5$). Scattering lengths $\bar{a}_{ij}$. a) Core width $d$ defined in the text. b) Density of ring component. c) Flow velocity $v_{1}$. d) Mach number $M_{1}$ of the ring component along the core.}
\label{flowvelos}
\end{figure}

As long as the features of the Skyrmion are sufficiently larger than the wavelength of the sound-modes in question, we can approximate the background two component BEC as locally homogeneous and apply the standard result for the speed of low energy excitations \cite{book:pethik}
\begin{align}
c_{\pm}&=\Bigg(\frac{1}{2m}\bigg(U_{11}n_{1} +U_{22}n_{2} 
\CR
&\pm \sqrt{ \left(U_{11}n_{1} -U_{22}n_{2}  \right)^{2}  + 4n_{1} n_{2} U^{2}_{12} } \bigg) \Bigg)^{\frac{1}{2}}.
\label{speedofsound}
\end{align}
Here $n_{1}$ and $n_{2}$ are densities of the ring and line components respectively. Demanding that both branches of the two-component speed of sound be real yields the condition $U^{2}_{12} < U_{11}U_{22}$, which is violated in the homogeneous Skyrmion system, \eref{phasesep}.  We therefore only consider the real branch of \eref{speedofsound}. We define the Mach number $M_{i}=v_{i}(\bv{x})/c_{+}(\bv{x})$, as the flow velocity of component $i$ expressed in units of the local two-component speed of sound. \fref{flowvelos} shows $M_{1}$ along the Skyrmion core, the core width (defined as $d(z)=\{ [ \int  (x^{2} + y^{2})|\Psi_{1} |^{2} dx dy ]/ [ \int  |\Psi_{1} |^{2} dx dy ]\}^{1/2}$ where the integration is over the inner core area only), its density, and the unrescaled flow speed for a typical Skyrmion close to the stability boundary.  The speed of the ring component though the thin vortex core of the line component has a sharp peak in the centre and reaches $M_{1} \sim 0.15$. 

\subsection{Offset of the two trap centres}
\label{offset}
As mentioned in \sref{model} and discussed in the Appendix, the hyperfine states of the two-component BEC will in general experience differential gravitational sag, and hence slightly offset traps. We numerically investigated this by introducing a trap offset in the $z$-direction $\mathbf{d}=d \hat{\mathbf{z}}$. We found for a Skyrmion far from the stability boundaries of \fref{lineofstability} b) ($N_{tot}=9 \times 10^{6}$ and $N_{r}=0.5$ \cite{savage:skyrm} with a radius of approximately $40$ microns) that an offset of $d=2 \mu$m produced instability. The hyperfine components spatially separate and the topological structure breaks down. However, with an offset of $d=0.1 \mu$m the Skyrmion remained stable. 

As the offset scales as $\omega_{z}^{-2}$ (see Appendix) it can always be made sufficiently small compared to the Skyrmion length scale $a_{z}\propto \omega_{z}^{-1/2}$ by using a tighter trap.  Skyrmions in tight traps also require lower atom numbers, by the scaling argument given in \sref{model}.
Another option is to operate at the magnetic bias field where the states experience the same trapping frequencies, which for our states is $B=3.23$ G \cite{cornell:scattlength}. In the Appendix we relate the trap offsets investigated above to magnetic field strengths in the vicinity of this value. Finally, it might be advantageous to create Skyrmions in a tight optical trap, which would also widen the range of suitable hyperfine states to include $ |F=2,m_{f}=2\rangle$ and $ |F=1,m_{f}=1\rangle$, between which exists an interspecies Feshbach resonance \cite{sengstock:mixedfr}.

%%%%

\section{Stabilisation against drift}
\label{drift}

Line vortices in single component trapped BECs are unstable towards drift out of the trap. However they may be stabilized by rotating the trap \cite{fetter:review,fetter:vortexstability}.  The Skyrmion line vortex behaves in the same way \cite{savage:skyrm}.

As there is no obvious experimental way to rotate the components independently, we rotate both. While the filled vortex in the line component is stabilized by rotation faster than a critical value $\Omega_{c,l}$, there is a destabilizing effect on the toroidal flow in the ring: for rotation frequencies $\Omega$ bigger than a critical value $\Omega_{c,r}$, the imaginary time evolution  creates an additional line vortex in the ring component. Due to the associated phase variation the topological equivalence of $\mathbb{R}^{3}$ and $S^{3}$ breaks down and with it the Skyrmion.

\begin{figure}
%\centering
\epsfig{file={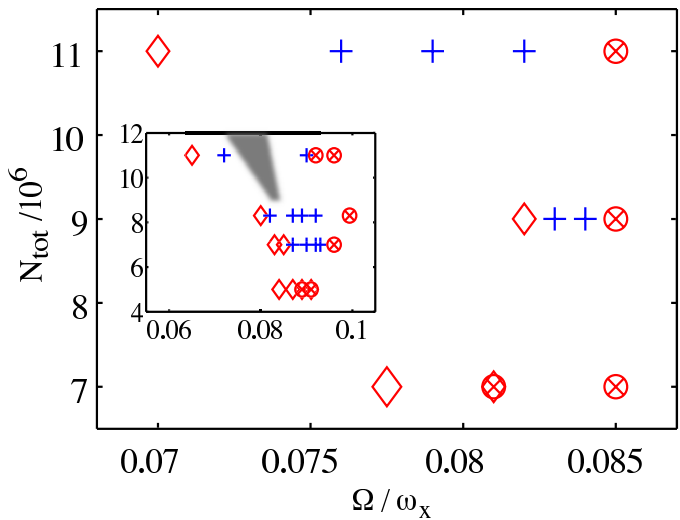},width=9cm}
\caption{(color online) Windows of stable rotation frequency for various total atom numbers; $N_{r}=0.5$ thoughout. 
\\
 ($+$ blue) stable, ($\varotimes$ red) unstable against additional vortex creation in the ring component, ($\diamondsuit$ red) unstable against line vortex drift out of the trap. 
Main diagram: $a_{ij}$ set of scattering lengths. Inset: $\bar{a}_{ij}$ set of scattering lengths. The grey shaded area indicates the location of the stability region for the $a_{ij}$ values.}
%($\times \! \! \! \! \! \diamondsuit$ red) unstable against line vortex drift out of the trap.}
\label{rotationwindow}
\end{figure}

Therefore stabilization requires $\Omega_{c,l}<\Omega<\Omega_{c,r}$. The critical frequencies $\Omega_{c,l}$ and $\Omega_{c,r}$ are determined numerically. 
To establish whether there is a stable Skyrmion for a given rotation frequency, we shift a previously obtained stable cylindrically symmetric state both parallel and perpendicular to the line vortex and evolve it in imaginary time. We observe either stabilization close to the original symmetry axis, or vortex drift out of the trap, or the disruption of the ring component circulation due to the creation of additional vortices. 
The response to trap rotation is studied for both sets of scattering lengths and varies significantly.
%Two different offsets were employed: (i) $\Delta y=2 \mu$m $\Delta z=2 \mu$m and (ii)  $\Delta y=10 \mu$m $\Delta z=10 \mu$m. Large perturbations might occur during Skyrmion initialization.
The resulting windows of stable rotation frequencies are shown in \fref{rotationwindow}, with the shape of the data points distinguishing the different behaviour in imaginary time outlined above. 
The critical rotation frequency for stabilization of the line vortex increases with decreasing atom number, as is the case for an unfilled line vortex in a pancake trap \cite{fetter:vortexstability}, while the critical rotation frequency for disturbances in the ring component decreases slightly for smaller numbers. Therefore the window that allows the stabilization of Skyrmions by rotation closes around $N_{tot}=8 \times 10^{6}$ for the early set of scattering lengths ($\bar{a}_{ij}$) and around $N_{tot}=9 \times 10^{6}$ for the revised ones ($a_{ij}$), see \fref{rotationwindow}. 

For low atom numbers, Skyrmions can be stabilized by a laser pinning potential \cite{raman:laserpinning} of the form \cite{laserbook}:
\begin{align}
%:
&V(\vc{x})=V_{0}  \left(\frac{w_{0}}{w(z)} \right)^{2}  \exp{ \left[ -2(x^2 +y^2)/w(z)^{2}  \right]} ,
\CR
&w(z)=w_{0}\sqrt{1+ \left( \frac{z}{z_{r}} \right)^{2}  } \; , \;
z_{r}=\pi \left( \frac{w_{0}}{\lambda} \right) w_{0},
 \label{laserpinning}
\end{align}
where $\lambda$ is the wavelength of the laser and $w_{0}$, $V_{0}$ are constants that describe its width at the focal point and intensity, respectively. Using the early scattering lengths, we found a stable  Skyrmion with $N_{tot}=2 \times 10^{6}$ and a pinning potential of $V_{0}=10 \hbar \omega_{x}$, $w_{0}=1.3\sigma_{x}$ ($\sigma_{x}\equiv (\hbar/m\omega_{x})^{1/2}$) and $\lambda=450$ nm.

\begin{figure}
\centering
\epsfig{file={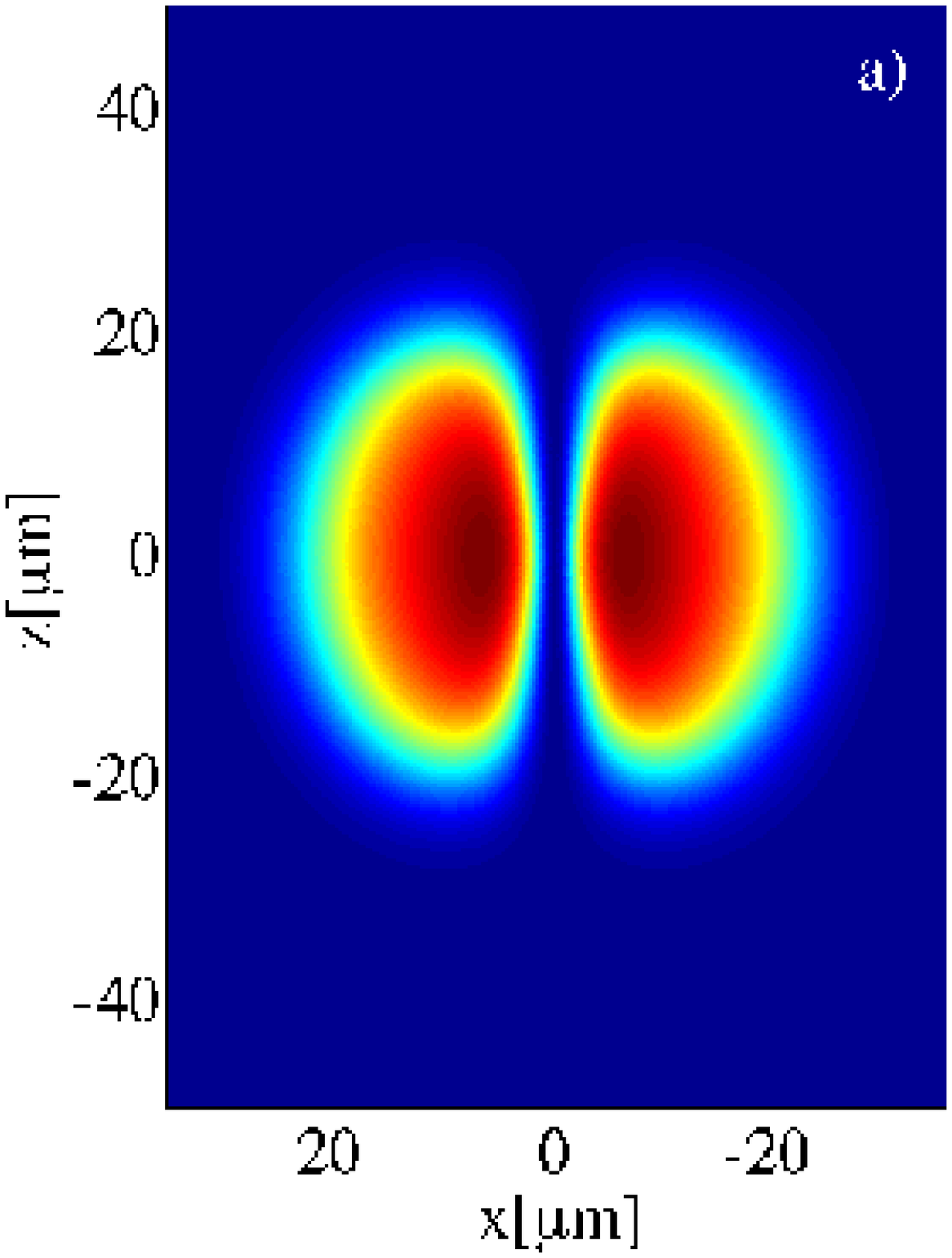},width=\smallfiguresize}
\epsfig{file={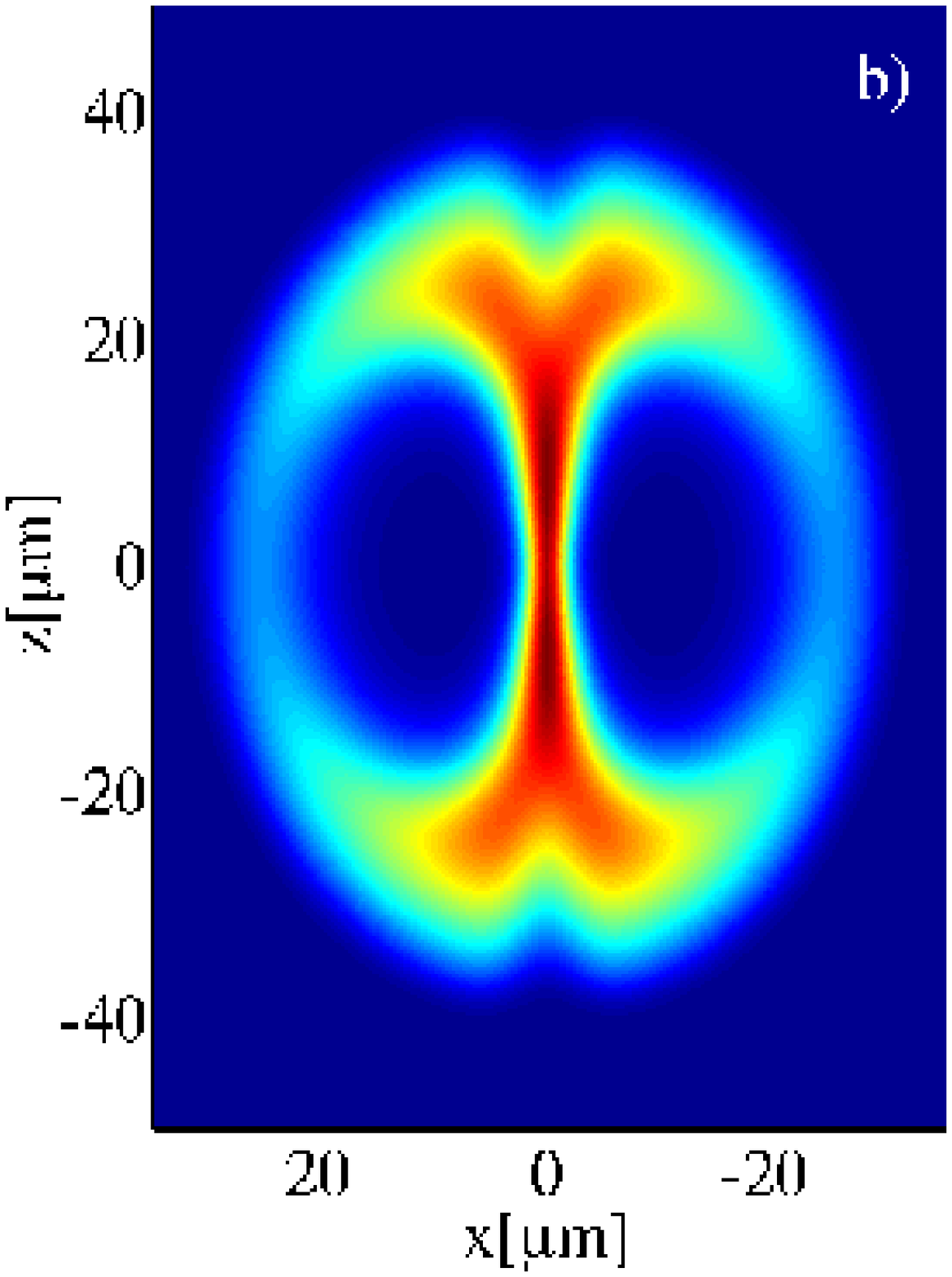},width=\smallfiguresize}
\\
\epsfig{file={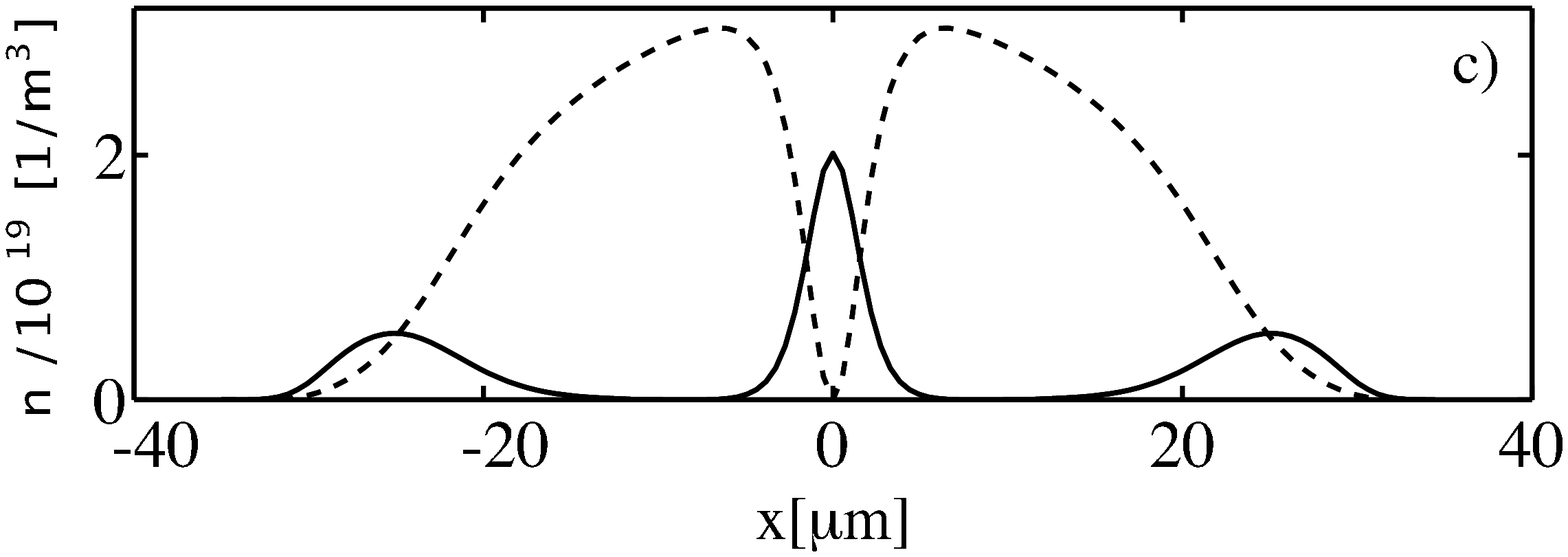},width=\figuresize}
\caption{(color online) Small Skyrmion with $N_{tot}=2 \times 10^{6}$ and $N_{r}=0.6$, stabilized against drift with rotation and pinning. Scattering lengths $\bar{a}_{ij}$. a) Colormap of line component density (blue - lowest density, red - highest density) in the x, z plane. b) Ring component density. c) Density profiles of the line (dashed) and ring components (solid) along the x-axis (y=0, z=0). Parameters as for \fref{3dplot}.}
\label{smallskyrm}
\end{figure}

However, the pinning laser beam can also have a destabilizing effect on the Skyrmions: it reduces the density of the ring component in the core, increasing the flow velocity, which can lead to collapse of the ring vortex. If laser pinning and rotation are used together, it appears to be possible to limit the disrupting effects of both methods. 
%In order to stabilize the Skyrmion with $N_{tot}=2 \times 10^{6}$ reported in the previous section, a laser pinning potential alone is not sufficient. The pinning laser beam reduces the density of the ring component in the core, increasing the flow velocity, and leading to collapse of the ring vortex. If laser pinning and rotation are used together, it is possible to limit the disrupting effects of both methods.
\fref{smallskyrm} shows an example, the same as in \fref{3dplot}, with the pinning amplitude reduced to  $V_{0}=6.5 \hbar \omega_{x}$, and trap rotation of $\Omega=0.06\omega_{x}$. 
The scattering lengths here are also the early values, $\bar{a}_{ij}$. We verified that this case is stable for independent perturbation of any one of the parameters mentioned above by $3\%$. 
The revised $a_{ij}$ scattering lengths also allow a small stable Skyrmion with combined rotation and pinning, but require $N_{tot}=2.13 \times 10^{6}$ atoms.
%We found that this Skyrmion could also be stabilized with a pinning potential of strength $V_{0}=10 \hbar \omega_{x}$ alone.

\begin{figure}
\centering
\epsfig{file={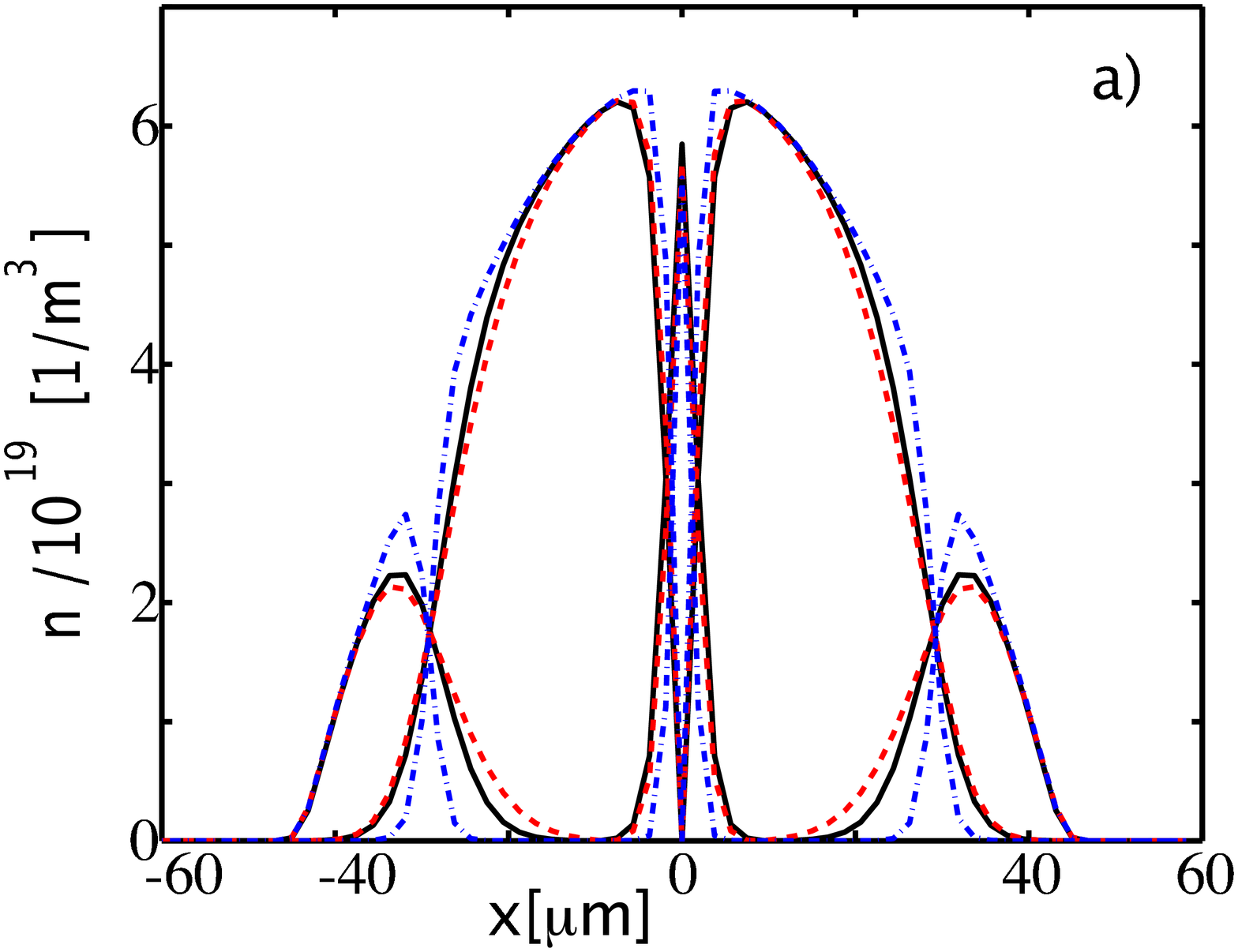},width=\figuresize}
\\
\epsfig{file={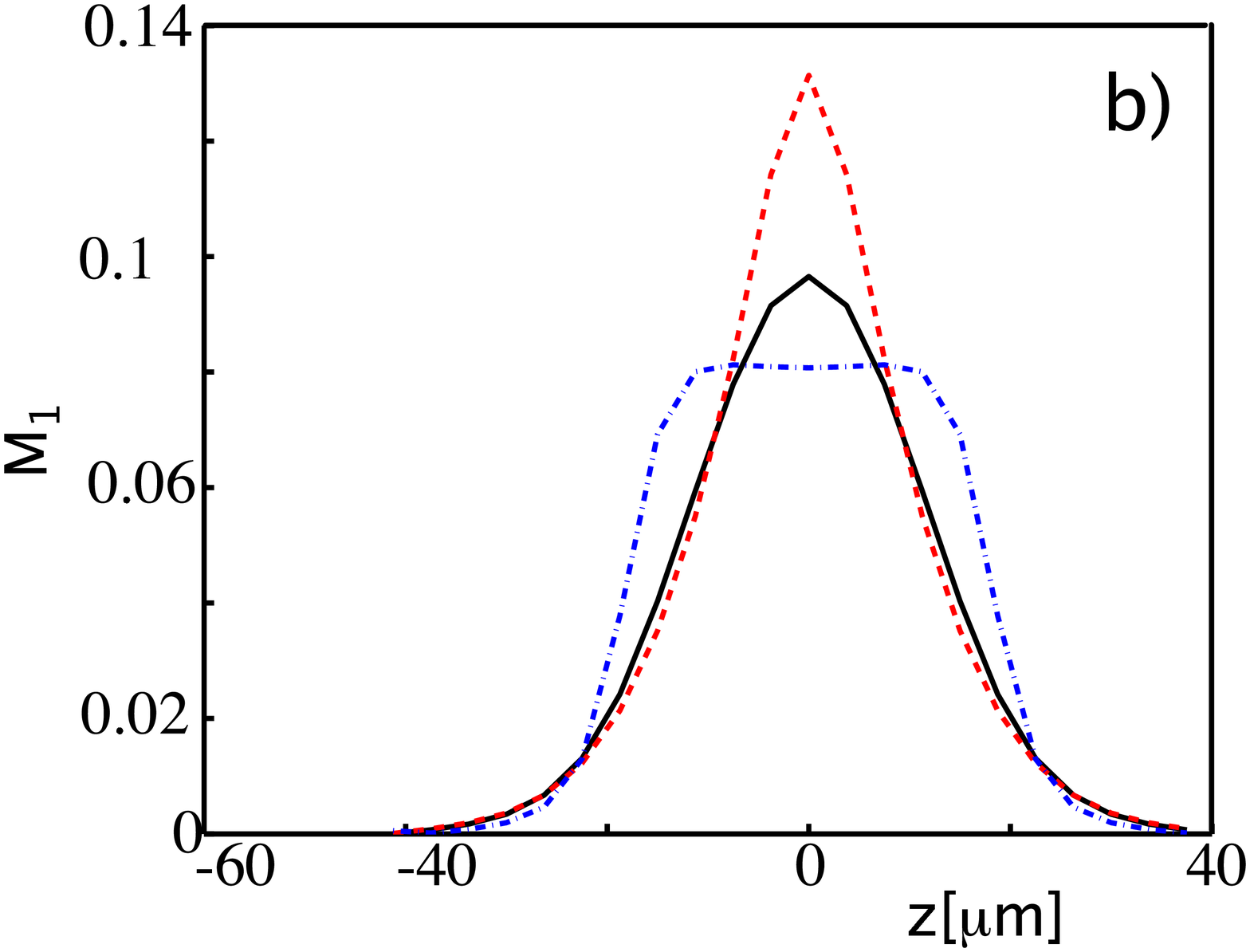},width=6cm}
%\\
%\epsfig{file={skyrm_paper_surftens3.eps},width=\smallestfiguresize}
%:
\caption{(color online) Skyrmion with $N_{tot}=9 \times10^{6}$ and $N_{r}=0.5$ for different surface tensions $\sigma$. 
\\(solid black) normal $\sigma$: $\bar{a}_{ij}$ set of scattering lengths. 
\\
(dashed red) low $\sigma$: $\bar{a}_{11}$ increased by $2\%$. 
\\
(dot-dashed blue) high $\sigma$: $\bar{a}_{12}$ increased by $5\%$. 
\\
a) Densities of the line- and ring components across the core (z=0, y=0). b) Mach number ($M_{1}=v_{1}/c_{+}$) of the ring component along the core (x=0, y=0).
}
\label{surftensplots}
\end{figure}

\section{Scattering lengths and Feshbach resonances}
\label{scattlength}
Due to potential uncertainties in the $^{87}$Rb (revised) scattering lengths, we investigated the effect of changing them on Skyrmion stability. For a case with 9 million atoms we found that variations  of $a_{11}$ by $5\%$ resulted in Skyrmion instability. For increased $a_{11}$ this was due to disruption of phase separation. For decreased $a_{11}$ the line component moved outward through the ring component. This is because the relative magnitude of $a_{11}$ and $a_{22}$ determines which component surrounds the other. In the Skyrmion configuration $a_{11}$ is larger and hence lower densities of the ring component are energetically favorable. We did not find that increases in $a_{12}$ were destabilizing. However, decreases of $2\%$ caused instability, again due to loss of phase-separation. Similarly, decreases in $a_{22}$ were benign, but increases of $2\%$ lead to instability. To verify some of these results we used numerical grids with $256^{3}$ points.

%Rescaling the new scattering lengths according to $\tilde{a}_{ij}=c a_{ij}$ with $c=0.937$ results in $\tilde{a}_{11}=5.67$, $\tilde{a}_{22}=5.39$ and $\tilde{a}_{12}=5.54$ and c can be absorbed into a minor rescaling of the atom numbers. Comparison with the older values $a_{ij}$ shows this situation to be equivalent to a $1\%$ increase in $a_{11}$ and a $0.7\%$ increase in $a_{12}$. Our results indicate that this lies within the stability domain, and simulations confirm that Skyrmions are stable with the $\bar{a}_{ij}$ scattering lengths, with almost unchanged surface tension compared to the $a_{ij}$. 

The sensitivity of Skyrmion stability to the scattering lengths can be understood in terms of the phase separation criterion or more generally in terms of the surface tension $\sigma$ at the component interface \cite{battye:phasesep,AoChui:surfacetens}. For an interface within a homogeneous background, and in the limit of $U_{11}=U_{22}$, \cite{AoChui:surfacetens} $\sigma$ is
\begin{align}
\sigma=n_{2}\sqrt{\frac{\hbar^{2}}{2m}\left(\frac{U_{12}}{\sqrt{U_{11}U_{22}}} -1 \right)\frac{U_{22}}{n_{1}}} .
\label{surftens_equation}
\end{align}
The derivative of this with respect to $U_{11}$ has a pole where $\sigma$ has a zero (the critical value for phase separation). It then predicts large variations in surface tension for small variations in $U_{11}$. 
Even though we find significant deviations from the homogeneous results, like stable Skyrmions in the regime $U_{12}^{2}<U_{11}U_{22}$, this qualitative prediction from the homogeneous case holds.
The zero of $\sigma$ is shifted from $U_{12}^{2}=U_{11}U_{22}$ in the trapped case, but our simulations show big changes in surface tension for small changes in the $U_{ij}$, as displayed in \fref{surftensplots}. 
\fref{surftensplots} a) illustrates the variation in the size of overlap regions between the line and ring components. \fref{surftensplots} b) shows the Mach number. The qualitative differences result from changes in the shape of the core.

\section{Conclusion}

Some useful conclusions may be drawn from our study regarding the experimental prospects for producing stable Skyrmions in trapped BECs. The situation is more optimistic than initial work might have suggested \cite{savage:skyrm}, as we have shown that substantially less atoms are required for stable Skyrmions to exist. However, we have also found that the windows of stability can be rather small, and that they are sensitive to the precise values of the scattering length ratios. The ability to manipulate the scattering lengths, for example using optical Feshbach resonances, would improve the robustness and flexibility of experiments. Stabilization techniques such as laser pinning are also likely to be useful.

\acknowledgments
S.W. would like to thank Beata J. D\c{a}browska for inspiring discussions and J. Hope for help with the computations. We also thank N. P. Robins, C. Figl and J. Close for helpful discussions. Finally we thank B.J. Verhaar for information regarding scattering lengths. This research was supported by the Australian Research Council Centre of Excellence for Quantum-Atom Optics and by an award under the Merit Allocation Scheme of the National Facility of the Australian Partnership for Advanced Computing. 

\begin{appendix}
\label{appendix}

\section{Application of the Breit-Rabi equation}

This appendix outlines how the magnetic field dependence of the magnetic moments of the hyperfine states constituting the Skyrmion leads to different trap frequencies. 
The energy of an atom in the hyperfine state $|F,m_{F}\rangle$, with electronic spin $1/2$, in an external magnetic field $B$ is given by the Breit-Rabi equation \cite{rabi,woodgate}
\begin{align}
E(F,m_{F},B) = & -\frac{\Delta_{hf}}{2(2I+1)} -g_{I}\mu_{B} m_{F}B 
\nonumber \\
&\pm \frac{\Delta_{hf}}{2}
\left(1 + \frac{2 m_{F}}{I+\frac{1}{2}} x +x^{2} \right)^{1/2} ,
\nonumber \\
x&=\frac{(g_{I}+g_{J})\mu_{B}}{\Delta_{hf}}B. 
\label{rabi}
\end{align}
The nuclear spin for $^{87}$Rb is $I=3/2$.
Here $\Delta_{hf}$ is the hyperfine splitting at zero magnetic field, $g_{I}$ and $g_{J}$ are the gyro magnetic ratios of electron and nucleus respectively and $\mu_{B}$ is the Bohr magneton. $x$ is a dimensionless magnetic field and the upper sign is for states with $F=I+1/2$, the lower one for states with $F=I-1/2$.
For $^{87}$Rb \cite{arimondo:review,salomon:values}
\begin{align}
\Delta_{hf}&=6834.68261090434(3) \mbox{MHz} , \nnl
g_{I}&=0.9951414(10) \times 10^{-3} , \\
g_{J}&=2.00233113(20).
\end{align}
The following discussion is one dimensional for simplicity. We 
assume the magnetic trap has the form
\begin{align}
B&=B_{0} +B_{2}x^{2},
\label{magtrap}
\end{align}
where $B_{0}$ is the bias field. We can expand Eq.~(\ref{rabi}) around some $B_{0}$ as
\begin{align}
E(i,B)&=E^{(i)}_{0}(B_{0}) +E^{(i)}_{1}(B_{0})(B-B_{0})  
\nnl
&+{E^{(i)}_{2}}(B_{0})(B-B_{0})^{2} + {\cal O}(B-B_{0})^{3},
\end{align}
where all information concerning the state has been gathered in the collective index $i$. 
Inserting the magnetic field, we obtain the following potential for atoms in state $i$
\begin{align}
V(x)^{(i)}&=E_{0}^{(i)}(B_{0})  +E_{1}^{(i)}(B_{0}) B_{2}x^{2}+{\cal O}(x^{4}).
\end{align}
The ratio of trap frequencies for atoms in states $i$ and $j$ at bias field $B_{0}$ is therefore
\begin{align}
\frac{\omega^{2}_{(i)}}{\omega^{2}_{(j)}}&=\frac{E_{1}^{(i)}(B_{0})}{E_{1}^{(j)}(B_{0})}
\end{align}
A gravitational acceleration $g$ will displace the equilibrium position in the trap from $x=0$ by 
$X^{(i)}_{0}={g}/{\omega^{2}_{(i)}}$.
The difference in the gravitational sag between the two hyperfine states is then
\begin{align}
\frac{|X^{(j)}_{0}- X^{(i)}_{0}|}{X^{(j)}_{0}}&=\left|1 -\frac{\omega^{2}_{(j)}}{\omega^{2}_{(i)}}\right|=\left|1-\frac{E_{1}^{(j)}(B_{0})}{E_{1}^{(i)}(B_{0})}\right| ,
\end{align}
If we apply this result to the $7.8$ Hz trap and hyperfine states considered in this paper, the resulting offset is about $11 \mu$m at a bias field of 1 Gauss. Eq.~(\ref{rabi}) predicts zero offset at $B=3.22946$ Gauss. The offset is about $2 \mu$m at $B=2.85$ Gauss and about $0.1 \mu$m at $B=3.21$ Gauss, which are the values considered in \sref{offset}.

\end{appendix}

\end{document}